# Abstract

**Method and apparatus for automatic text input insertion in digital devices with a restricted number of keys.**

**Nikolaos Tselios and Manolis Maragoudakis**

A device which contains number of symbol input keys, where the number of available keys is less than the number of symbols of an alphabet of any given language, screen, and dynamic reordering table of the symbols which are mapped onto those keys, according to a disambiguation method based on the previously entered symbols.

The devices incorporates a previously entered keystrokes tracking mechanism, and the key selected by the user detector, as well as a mechanism to select the dynamic symbol reordering mapped onto this key according to the information contained to the reordering table.

The reordering table occurs from a disambiguation method which reorders the symbol appearance. The reordering information occurs from Bayesian Belief network construction and training from text corpora of the specific language.

# METHOD AND APPARATUS OF AUTOMATIC TEXT INPUT IN DIGITAL DEVICES WITH A REDUCED NUMBER OF KEYS.

## Description of the invention

### Technical field of the invention

The proposed invention deals with the data entry usability issue in digital hardware where the number of available keys is smaller than the number of symbols (alphabet) of any given language. More specifically, the invention introduces a novel automatic text entry method in digital devices where the keyboard has two or more letters mapped onto each key. Such a device is, for example, a telephone handset (cellular or conventional phones), pocket calculators, palm and pocket pc's and remote controllers.

### Evaluation of current state-of-the-art

In a mobile phone, the letters of an alphabet have to be mapped onto a nine-keypad. As a consequence, this means that more than two letters have to be grouped in one single key. Due to that reason, usually more than one keystroke is required in order for a user to access and enter a letter, while he writes a text message.
Nowadays, three alternative dialogues have been established in order to assist the user in editing a message.

A. The 'conventional', multiple key pressing symbol entry method, to enter a given symbol.

The simpler, yet widely acceptable, which from now will be referred as STEM (Standard Text Entry Method), approach requires tapping the corresponding key as many times as needed to appear on screen for a letter to be entered. The basic disadvantage of multiple keystrokes is the slow rate of the inserted letters. However, as previously described, this lack of speed influences positively the need for user confirmation. So, the user does not have to pay any attention to the mobile phone screen. Another problem appears when typing two letters that lie in the same key. The most common solution is the introduction of a time delay (timeout) between two taps of the same key, in order to verify that the user wants to type two letters from the same group or one letter by multiple taps. As an alternative, in order to bypass the timeout delay, the user can depict his/her intention to type two letters from the same key by selecting a certain timeout kill key which cancels the timeout delay. This obviously further deteriorates the message editing speed. Additionally to the poor task execution time provided by this method, extensive effort in terms of keystrokes is required from the user in order to complete typing a message. Task execution time is further increased in general with the introduction of the timeout delay time.

B. Symbol entry method using two keys.

Another, similar approach is the two-key input method, in which a user specifies a character by pressing two keys. The first key represents the group of letters (e.g key 2 for A, B or C) and the second disambiguates the letter by selecting its place in the group (e.g key 1 would select A). Studies (Silfverberg et. al., 2001) have depicted that although the two-key approach is very simple, it is not efficient for Roman characters, since there is great loss of speed by moving between the two keys. This result could be generalized in a manner that the ascertainment of the low efficiency of the method remains valid in alphabets other than Roman. That is probably the main reason why this method is not popular among users. Note however that it is very common for typing Katakana characters.

C. Dictionary based symbol entry methods.

Another category of methods developed to deal with the disambiguation problem, uses a dictionary in order to deal with letter disambiguation. Among the lexicon-based methods, the most popular is called T9©, developed by Tegic©. The T9 is fully disclosed in patent US5818437. More specifically, the user presses the key in which the desired letter lies, only once. While we are pressing some letters or by the time a word is completed, which means that a space was entered, the system is trying to output the most probable word that corresponds to the key sequence that the user provided. If the guessed word is incorrect, then using a special key the system outputs a pool of other words that also correspond to the specific key sequence.

This method significantly reduces editing speed but requires user attention and since it is based on a lexicon, it cannot efficiently handle unknown or shortened words, slang, names etc., heavily used in mobile text messaging (Longmate et. al 2001). Another important drawback of T9 is the poor feedback to the user during the process of typing a word. There are times that letter disambiguation occurs at the latter characters of a word, so until then, the user may see a totally different set of characters, a phenomenon that obviously results in user confusion due to reduced sense of progress towards user's text entry goal.

Having discussed the advantages and the disadvantages of STEM (and it's two key variance) and T9, the aim of the presenting method and apparatus is to combine the strong points of the above-mentioned methods (high learnability, predictable dialogue flow for the STEM process, increased efficiency for the T9 method) in a brand-new novel approach, without the disadvantages they have (poor performance, confusing feedback to the user). The proposed invention contributes to the current state of the art, given the fact that is not characterized by the disadvantages of lexicon based methods, where the inputs to be given affect the result of previous given keystrokes. Additionally, it enables symbol selection with only a key press at a time, independently from the position they posses into they key where they are mapped to.

Thus, the advantages in brief of the present invention which in total promote the state of the art in the digital appliances where text entry is possible with a limited number of keys are the following:

- The required effort in terms of keystrokes required, in digital appliances where the number of available keys of keyboard is smaller than the number of symbols of the alphabet of any given language, is decreased considerably.
- It is not based on a dictionary to handle the disambiguation problem. As a result, it can handle unknown or abbreviated words more effectively, simplified idioms and slang language, the names etc.
- The information that alters the ambiguities and dynamically reorders the letters in the keys of such an appliance is constant and limited size. Thus, it is easy to simultaneously incorporate of many such structures of information (e.g. different languages), more that the corresponding methods that are based on dictionary.

- It is an innovative method that is not characterized by the disadvantages of lexicon based methods, where sequences of forthcoming tapings influence the result of previous given tapings, thus providing constant feedback to the user.
- It is characterized by high interaction learning rate (learnability), because a single typing is required for the desirable letter to appear. In case where the presented letter is not the desired one, the user uses a specific key that presents the second most likely letter and so on. Afterwards, he continues message entry task.

Brief description of the method.

In various digital devices such as mobile and conventional telephone units, where the functionality of text input is supported but the physical dimensions of the telephone keypad are limited, three or more letters of an alphabet should be grouped together within one key. Due to that fact, more than one keystroke is usually expected from the user, in order to insert a letter when editing a short text message. (In other words, the user has to manually resolve the ambiguity of the desired letter that relies on a group of letters).

The proposed method automatically resolves this ambiguity by allowing the user to select the desired letter, given the pressed key and the immediate sequence of letters that precede the desired.

The result is obtained by a database that reorganizes the priority of the letter to appear, given the above mentioned information. The reorganizing information is pre-stored in the device and the framework to obtain it is based on a Bayesian Belief network (BBN) structure. This structure offers a machinery of estimating the user's desired letter given the letters that precede. BBN are used in the machine learning community as a means of effective reasoning under uncertainty, they are able to learn to predict the state of a variable given the states of all, or a subset of other variables from the available data. BBN appear to behave very successful in the proposed algorithm, due to the restrictions posed by this algorithm. Since the letters are usually grouped into classes of three or four letters (this number is related to the number of letters an alphabet has), a user desires to input one of them (and not one of the whole alphabet). The network is expected to estimate a letter from the limited candidate letters, given the letters that precede in the current word, this estimation is a very good approximation of the real case.

The primary aims of this invention are:

•       To minimize the average number of key presses, in order to edit a text message.
•       To provide a device and a method that will achieve the above for any language.
•       To provide consistent and constant feedback to the user, in terms that during a text message editing, the depicted symbols will not be modified but only if a user explicitly declares it to the system.
•       To reduce the expected number of keystrokes as opposed to the STEM method, using stored information that dynamically reorders the layout of letters within a key. This information is extracted from a machine learning method and affects the letter layout based on the given keystrokes until the specific time
•       To maintain a small amount of memory load for storing that information, in order to incorporate the proposed method in any device, regardless of their memory capacity.
•       To provide a method for the optimal exploitation of memory in such a device.

The aims of the proposed methodology will analytically be discussed in the following text.

## Detailed description of the invention

In **Fig. 1**, a schematic representation of a typical telephone apparatus is provided, along with the corresponding letter organization within the keypad keys. The device that incorporates the presented method does not differ externally from any other devices that can be found in the modern market. No letters are grouped within keys 1 and 0, while in the rest keys (2-9) three or four letters are grouped, regarding the considered alphabet. For instance, the Greek alphabet contains 24 letters, thus three of them have to be grouped in the eight keys (24/8=3). Regarding the Latin alphabet, the two more letters (26 instead of 24) have to be mapped into two keys in groups of four.

As an example, in order to edit the Greek word "HMEPA" (day) with the conventional method, one should press key 4 once in order to select the letter "H", three times key 5 for the letter "M", twice key 2 for letter "E", twice key 7 for letter "P" and once key 2 for letter "A". The total number of key presses is 9, with an average of 9/5=1,8 keystrokes per letter. The aim of this invention is to bring this average in a scale of 1 keystroke per letter.

If we suppose that the letter layout was: for key 5, Μ-Λ-Κ, for key 2, Ε-Δ-Ζ and for key 7 Ρ-Π-Σ, then the number of the required keystrokes would be only 5 that is one keystroke per letter. The goal of this invention is to dynamically reorder the letter layout of a given keypad, so that the required key presses are equal to the number of the letters of a text message. By default, this in not always possible. For example, regarding the editing of the word "ΣΤΙΓΜΑ" and the word "ΣΤΗΡΙΓΜΑ" in the third keystroke, a different letter is expected from key 4, so that in one of the two words, be more than one keystrokes would unavoidable be needed.

The invention presents a method and an apparatus that achieves to significantly approximate the ideal case of one keystroke per letter. This method impels the letter layout of a key to be changed in any moment of the text input process, so that the expected keystrokes are minimized, and the apparatus operates, based on this method.

In **Fig. 2**, the way this device operates is depicted. Initially, the user selects the key in which the letter of his desire lies. (1). Afterwards, the device activates a method of dynamically modifying the letter layout, in order to better estimate the desired letter, given the letters than already precede in the text (2). This method is based on a machine learning method that learns from data, named as Bayesian Belief networks, presented in details in the following sections. In a following stage, the apparatus provides an optical or acoustic feedback by introducing the first letter that lies in the new letter layout that method made (3). The user observes this feedback (4). In case this is the desired letter, he continues the process of text editing by returning to step (1). If this is not the case, then he uses a key of selecting the next letter of the dynamic letter layout. This special key is chosen to be the dash key (#) and can be found in the lower rightmost part of the keypad. (5) By using the (#) key, the next letter of the dynamic letter layout appears. (6). Step (5) is repeated until the desired letter appears on the screen. Subsequently, a user continues the text editing process, returning to step (1).

In **Fig. 3**, the basic functionalities of this device are presented. The insertion of letters is achieved trough keys, such as those that conventional telephone devices incorporate. Signals from this key pad are transferred to a symbol layout selector that provides the layout of the symbols that are to be mapped in the key that was pressed. This is achieved, based on the key that has been pressed, the information the selector obtains from a temporal memory that contains the most recent keystrokes ( the most recent symbols that appear on the screen) and the final feedback that is provided by the dynamic modification of the letter layout. The symbol then appears on screen, while the new, dynamic layout is stored in a temporal memory. The number of the most recent letters that can be used from the device can very, however three preceding letters are considered to be the optimal number, taking the memory needs into consideration. Nevertheless, there are certain cases where a different number might be necessary.

The moment a user observes the symbol that appeared on the screen and discovers that this was not the one he desired, he presses the (#) key, which selects the next letter of the letter layout. Alternatively, the selection of another letter could be achieved by multiple presses of the same key that contains this letter, such as in the STEM method. In order to do this, a time delay slot (timeout) between two subsequent pressed of the same key is required in order the user does not select an additional letter from the same group.

The given symbol can be selected by continuously pressing of the key that contains it, in case that a symbol reordering is not required. For that reason, the information regarding the new layout of the letters is maintained until the user presses another key. Then, a related signal is lead to the selector, along with the last symbol that is currently appearing on screen. The symbol selector is estimating the symbol that follows in the reordering layout and has been stored in the temporal memory. Subsequently, the selector submits the necessary signal in order to appear on the screen, by confuting the letter that previously appeared on the screen. Note that the device may not necessarily incorporate a screen as a means of user feedback, but other means such as digital speech synthesis.

The method that is being described in order to infer on the probability of a letter that lies within a group of letters and the most recent letter sequence is based on BBNs. The proposed method uses this a-priori probability of a letter to be the desired one.

The principal idea of BBNs is that an inference problem can be considered as a domain where nodes are connected with arcs forming a directed acyclic graph, where the arcs define a kind of node relationships. Each node corresponds to a problem variable or the unknown and uncertain quantity. The variables can take discrete, finite states. The degree of node relationships is determined by a probability distribution that is based on the foundations of Bayes' theorem. As appearing on the related bibliography, BBN present significant advantages than other machine learning algorithms, such as neural networks, decision trees, etc. give the fact that is at least as efficient as the other methods and is based on a solid mathematical background (Stephenson, 2000).

In order to give a more precise mathematical definition, Bayesian Belief Network (BBN) is a significant knowledge representation and reasoning tool, under conditions of uncertainty. Given a set of variables D = <$X_1$, $X_2$…$X_N$>, where each variable $X_i$ could take values from a set T($X_i$), a BBN describes the probability distribution over this set of variables. We use capital letters as X,Y to denote variables and lower case letters as x,y to denote values taken by these variables. Formally, a BBN is an annotated directed acyclic graph (DAG) that encodes a joint probability distribution. We denote a network B as a pair B=<G,Θ>, (Pearl, 1988) where G is a DAG whose nodes symbolize the variables of D, and Θ refers to the set of parameters that quantifies the network. G embeds the following conditional independence assumption: *Each variable Xi is independent of its non-descendants given its parents.*

Θ includes information about the probability distribution of a value $x_i$ of a variable $X_i$, given the values of its immediate predecessors. The unique joint probability distribution over <$X_1$, $X_2$…$X_N$> that a network *B* describes can be computed using equation (1.1):

$$P_B(X_1...X_N) = \prod_{i=1}^{N} P(x_i \mid parents(X_i)) \quad (1.1)$$

In the process of efficiently detecting the optimal symbol reordering, concerning a given number of symbols entered, the BBN should be learned from training data (text corpus of the given language) provided. Learning a BBN unifies two processes: learning the graphical structure and learning the parameters Θ for that structure. In order to seek out the optimal parameters for a given corpus of complete data, we directly use the empirical conditional frequencies extracted from the data (Cooper and Herskovits, 1992). The selection of the variables that will constitute the data set is of great significance, since the number of possible networks that could describe these variables equals to $2^{\frac{N(N-1)}{2}}$, where N is the number of variables (the number of previous symbols entered taken into consideration).

We use the following equation along with Bayes theorem (well known in mathematics) to determine the relation r (or Bayes factor) of two candidate networks $B_1$ and $B_2$ respectively:

$$r = \frac{P(B_1|D)}{P(B_2|D)} \quad (1.2) \qquad P(B|D) = \frac{P(D|B)P(B)}{P(D)} \quad (1.3)$$

where:

- $P(B|D)$ is the probability of a network B to be the desired given data D.
- $P(D|B)$ is the probability the network gives to data D.
- $P(D)$ is the 'general' probability of data.
- $P(B)$ is the probability of the network before seen the data.

We apply the equation (1.3) to (1.2). Having not seen the data, no prior knowledge is obtainable and thus no straightforward method of computing $P(B_1)$ and $P(B_2)$ is feasible. A common way to deal with this is to assume that every network has the same prior probability with all the others, so equation (1.2) becomes (1.4):

$$r = \frac{P(D|B_1)}{P(D|B_2)} \quad (1.4)$$

The probability the model gives to the data can be extracted using the following formula (Glymour and Cooper, 1999, 1.5):

$$P(D|B) = \prod_{i=1}^{n} \prod_{j=1}^{q_i} \frac{\Gamma(\frac{\Xi}{q_i})}{\Gamma(\frac{\Xi}{q_i} + N_{ij})} \prod_{k=1}^{r_i} \frac{\Gamma(\frac{\Xi}{r_i q_i} + N_{ijk})}{\Gamma(\frac{\Xi}{r_i q_i})} \quad (1.5)$$

In equation (1.5) all the terms are known and computable. More specifically:

- $\Gamma$ is the gamma function.
- $n$ equals to the number of network variables.
- $r_i$ denotes the number of values (different states) in *i:th* variable.
- $q_i$ denotes the number of possible different value combinations the parent variables can take.

- $N_{ij}$ depicts the number of rows in data that have *j:th* value combinations for parents of *i:th* variable.
- $N_{ijk}$ corresponds to the number of rows that have *k:th* value for the *i:th* variable and which also have *j:th* value combinations for parents of *i:th* variable.
- $\Xi$ is the equivalent sample size, a parameter that determines how readily we change our beliefs about the quantitative nature of dependencies when we see the data. In our study, we follow a simple choice inspired by Jeffreys (1939) prior. $\Xi$ equals to the average number of values variables have, divided by 2.

By combining equations (1.4) and (1.5) we have an equation with all the terms known and computable. Therefore the realization of a function which calculates the optimal network structure which model of finding the most likely letter given a certain amount of previous keystrokes is feasible.

The level of complexity of BBN produced is increased depending on the number of previous entered letters that takes into consideration. Nevertheless, because the restrictions of memory of a digital appliance such as a mobile telephone, we do not examine the prefixes that are constituted from more by three letters.

At the same time, as it resulted from certain measurements, the effectiveness of the method is improved slightly, if we take into consideration more than three symbols that have proceeded. In case where the system forecasts inaccurately a letter, the proposed interaction dialogue gives the possibility of a special key (the dash key - #) to be used by the user that can change the output to the second more likely letter and so on.

In **Fig. 4**, the BBN structure is depicted. The network is considering three preceding letters as well as the key that was pressed, in order to estimate the desired letter. Node 'three letters before', 'two letters before' και 'one letter before' correspond to each prefixes. Node Key corresponds to the key that was pressed and takes as states the values of two to nine. Finally, node State has tree discrete values, namely one, two and three that represent the position of the letter in a letter layout. The network embeds a conditional probability table that can be consulted in order to predict which is the most probable position of a letter (one two or three), by considering all the other nodes of the network or a subset of them. For example, suppose that a user wishes to enter the work "ΗΛΙΚΙΑ" (age). Suppose also that the system has already predicted the part "ΗΛΙΚ" correctly. In order to insert the letter "Ι", the user presses key 4 where the letters "Η",

"Θ" and "Ι" are contained. The Bayesian network is about to predict the most probable letter, given the letter sequence of "ΛΙΚ" and the key 4 that was pressed. The most probable letter is inferred. In case this in not the desired one, by sequential presses of the (#) key, the system can change its prediction to the second or the third letter, according to the user desire. The evaluation of the effectiveness of the network was performed to a text corpus that was not used for training and was estimated to 95,5%.

The novel methodology combines the increase of text input speed with the dialogue consistency, particularly for words not found in a lexicon. Furthermore, the proposed method demonstrates low memory needs (only 10Kbytes or less for each supported language), a significant advantage when confronting with T9, a method that incorporates a lexicon of approximately 5000 words.

Indeed, for each language only the memory entry of a table with all the three preceding letter combinations is required. For example, regarding the Greek language, that is 24*24*24=13824 entries. Moreover, in each of those entries an 8-digit number is included. For each of the 8 keys (2 to 9) that the letters are grouped, a digit that describes the letter layout is mapped. So, for key 2, if the letter layout of the letters Α-Β-Γ is 1-2-3, this sequence can be symbolized with 1. In a similar manner, regarding 1-3-2 the digit 2 is used, for 2-1-3 the digit 3, for 2-3-1 the digit 4, for 3-1-2 the digit 5 και for 3-2-1 the digit 6.

In case the method utilized alphabets that contain more than 24 letters, more than three letters have to be mapped within a single key. So, in an alphabet of 26 letters (English), two keys will contain four letters. As a result, the combinations of those 4 letters is now 24 instead of 6, a fact that suggests that for those keys, two digits will be required in order to encode the possible letter layouts. Rather than having an 8-digit number, in cases of 26 –letter alphabets the method needs a 10-digit number. It is more than obvious that the method can be incorporated into any digital device with minimum cost.

## Evaluation of the method

As discussed in the following sections, a usage of a simplified version of the GOMS methodology and the KLM method (Keystroke Level Model – Model of Level of Typings) were utilized to compare of standard text entry method (STEM) and the new method which guesses automatically the desirable letter, showed that even a relatively mediocre precision in the estimate of desirable letter, leads to significant improvement of efficiency to the text messaging task.

Keystroke Level Model (KLM) is an analytical predictive method inspired by the Human Motor Processor Model (Card et al. 1980). This model focuses on unit tasks within a user-machine interaction environment which consists of a small number of sequenced operations. The model assumes two phases in task execution. During the first phase decisions are made on how to accomplish the task using the primitives of the system. During the second phase the execution of the task takes place without high level mental activity. The model assumes expertise from user and does not focus on the user interface interaction learning process. This method has been empirically validated against a range of systems and a wide selection of tasks, and the predictions made were found to be remarkably accurate (prediction errors less than 20%, Olson and Olson, 1990).

In our effort to evaluate the proposed method and apparatus, we assume negligible times for the system (mobile device) response and the mental operators (the user is assumed to have decided what to write and knows exactly the positioning of letters on the keypad), we can develop a model to predict times for an expert user to enter a word. According to this model the time to complete entry of a word using STEM is:

$$T_{STEM} = X[nT_P + T_{PER} + (1-P_{CK})T_{WAIT}] + (X-1)P_{CK}T_{CK} \quad (1.6)$$

And time to complete entry of a word using the proposed method (from now on called iPRETI intelligent PREdictive Text Input) is:

$$T_{IPRETI} = X[T_P + T_{PER}] + (X-1)P_{CK}T_{CK} + X(P_{ERROR1} + P_{ERROR2})(T_{CK} + T_P) \quad (1.7)$$

where:
- $X$ denotes the number of letters for a specific word.
- $n$ denotes the average number of keystrokes to select a specific letter using STEM (calculated 2.0229 from a sample of 386870 letters during greek word typing. Corresponding are the values for other languages).
- $T_P$ denotes average key press time. (165 milliseconds (Silfverberg et. al 2000)).
- $T_{PER}$ denotes time required from user to perceive correct entry. (500 milliseconds).
- $P_{CK}$ probability of requiring a letter contained in a different key than the previously pressed. (calculated 0.89 from a sample of 386870 letters).

- $T_{WAIT}$ time waiting for cursor to proceed, when successive letter contained in the same key.(depends on phone, for Nokia models is 1500 milliseconds (Silfverberg et. al 2000)).
- $T_{CK}$ required for a user to move to another key. (approximately calculated by using Fitt's Law: 215 milliseconds (Silfverberg et. al 2000)).
- $P_{ERROR1}$, $P_{ERROR2}$ are probability for a proposed letter not be the required one, and probability for the second proposed letter not be the required one, respectively. (calculated as 0.045 and 0.002 respectively).

Applying equations (1.6) and (1.7), we obtain $T_{STEM}$ = 5695,8 msec and $T_{iPRETI}$= 3590,5 msec for an average Greek word length (X=6). Increase in task efficiency is 34,72% in terms of time required and average number of keystrokes required is 12,13 and 6,39 respectively, a difference of 47,35%. Modeling of T9 method does not give accurate results because of the inconsistent behaviour of the algorithm. More specifically, the keystrokes per letter required is reduced to one, except from the cases where the first proposed word is not the one that user wants to enter forcing him to choose across a list of proposed words. Secondly, if the word required is not in T9's dictionary, the user has to alter the text entry method to STEM thus further reducing efficiency of the task. Unfortunately, no published study exists concerning the proportion of desired words present in the dictionary –especially for the Greek language, and on how often a word other than the desired one appears. Therefore, no accurate dialogue modelling can take place.

.

## Experimental evaluation of the method's efficiency

Having already theoretically modeled each technique's dialogue performance concerning the time to complete word entry, we intended to verify iPRETI performance in the real world. For that reason, we have implemented a mobile phone keypad emulator where users were supposed to edit messages using iPRETI.

The arrangement of the Greek letters in every key was identical to that of the current mobile phones available in the market. For our experiments, we considered only capital letters, since they are most commonly used by the Greek users.

This does not influence the attribution of method, since the alternation of capital/small letters becomes with the use of different key. Moreover, in the lower part, the system outputs the probability for each state of the last pressed key. Emulator traces the number of keystrokes using iPRETI and compares to those that would be needed by STEM for the same message. The right part of the simulator contains the graphical representation of the number of keystrokes needed by, during the editing procedure. This graph is dynamically updated across the editing progress, thus providing a better sense of each method's behavior.

As we could observe from an example text messaging task, iPRETI is better than STEM throughout the whole editing process with an average keystroke number that approximates 1.06. On the other hand, STEM converges to a value of about 1.94 which agrees to our initial expectations. Performance measurements in terms of time required to complete text entry task could not be compared directly to the KLM model at the moment, because of the non negligible response time required by the system to find the appropriate probabilities due to early prototyping issues.

To evaluate real world performance of the proposed method, we have conducted preliminary experiments using ten SMS prototype phrases of varying length containing high informal word rate. **Figure 5**, depicts the ten SMS phrases selected from actual written and submitted messages. **Figure 6**, tabulates analytic results concerning the number of keystrokes needed from iPRETI and STEM and error rates of single errors and double errors (e.g. second and third keystroke required to access desired letter respectively).

Having analyzed the results we could clearly distinguish an improvement of 37.4% concerning the effort required to edit a message in terms of keystroke number. Improvement in efficiency of interaction is one of the three core issues related to the usability of a user interface according to IS 9241. The percentage of correctly predicting a letter by iPRETI is 91.2%., considered very high despite the high percentage of informal words. Note that the average keystroke numbers excluding spaces within words for iPRETI and STEM are 1.118 and 1.907 respectively, depicting an improvement of 41.3%. A notable remark is that the extracted results have a close convergence to our initial predictions derived by KLM modeling. The results obtained clearly confirm the accuracy of the disclosed invention (Maragoudakis, Tselios, Fakotakis, & Avouris, 2002).

# Claims

1. An embedded device that consists of input symbol keys where the number of available keys is less than the number of the symbols of an alphabet. At least 3 keys do not contain any mapped symbols. The device also contains a screen in which various information is depicted, including text messages during the process of editing, sending, receiving and manipulating them in general.

2. The device of claim 1 embeds a database of reordered symbols that are grouped in one of those keys, based on a disambiguation method. The database is used in order to select the desired letter when pressing a key, provided the preceding symbols.

3. The device of claim 1 embeds a pool of mechanisms that record the most recent sequence of key presses, as well as the identity of the key that contains the desired letter. An additional mechanism selects and reorders the symbols based on the stored information which lies in the reordering table and in the record of the most recent key presses.

4. The device of claim 1 embeds a mechanism of transmitting and revealing of the pressed symbols on screen, as well as a mechanism of informing the user on the style of text editing (either the proposed one or the traditional method of multiple presses per key).

5. The device of claim 1 incorporates the symbol disambiguation method, which is characterized by the fact the it uses information that reorders the priority of certain symbols to appear. This method is based on learning Bayesian Belief networks from data obtained by text corpora of a given language. The necessary information in store in a table of at most $(y+1)^n$ rows and 2 columns, where $y$ denotes the number of symbols an alphabet contains and $n$ is the number of letters that a Bayesian network take into consideration in order to infer on the desired letter. The method obtains input by the mechanism which records the most recent presses and provides a specially designed information as regards to the new order the symbols should appear to the user, for every keys that the device of claim 1 contains.

6. The reordering table, described in claim 5, all the possible combinations (depending on the number of the recently appearing symbols) are stored. The optimal number of recent symbols in terms of computational efficiency is 3. However, this does not restrict the case one might desire to consider a greater or a lower number of symbols, in order to build the reordering table, if that is considered as necessary.

7. The device of claim 1, based on claims 2 and 5 is characterized by the fact that it incorporates as many reordering tables as the number of languages it can support, in the process of helping the text editing procedure.

8. The reordering database that the device of claim 1 uses and is being analytically described in claims 2,5,6,7 does not require to be physically embedded to the device. It can be called through the telephone connection from the corresponding mechanism of claim 3.

9. The device of claim 1 is characterized by the fact that it uses additional keys in which no symbols are mapped. Such keys enable or disable the method as well as the language of the message.

10. The device of claim 1 is characterized by the fact that is uses an additional key in which no symbols are mapped in order to change the symbols of a language to uppercase or lowercase.

11. The device of claim 1 is characterized by the fact that it uses an additional key in which no symbols are mapped in order to successively select the next most possible desired letter, in case that the current shown according to the dynamic reordering process is not the desired one. This key could be the dash key (#) due to the fact that it is embedded in every mobile device nowadays, therefore the user is used to its presence, and its rarely used for another scope.

12. The disambiguation method is characterized by the fact that it leads to a significantly reduced number of keystrokes to devices such as those described in claim 1, where the number of available keys is less than the symbols of any given language's alphabet.

13. The disambiguation method typically has a successive (desired letter selection) rate above than 90%.

14. The disambiguation method according to claims 1 and 2, is independent from any embedded lexicon and is based upon to letter reordering table according to the preceding keystrokes.

15. The disambiguation method according to claims 1,2,5,6,7 is achieved with altering the ambiguities in a letter-symbol level and not in a word level of the language where is applied.
.

16. The disambiguation method according to claims 5 and 6 is independent from any other symbols independent from the symbols of the given alphabet. According to this, for example the keystroke sequence Α.Π (the intermediate symbol is the point symbol) is considered for the method as Α_Π (where the underscore (_) symbol represent a single space).

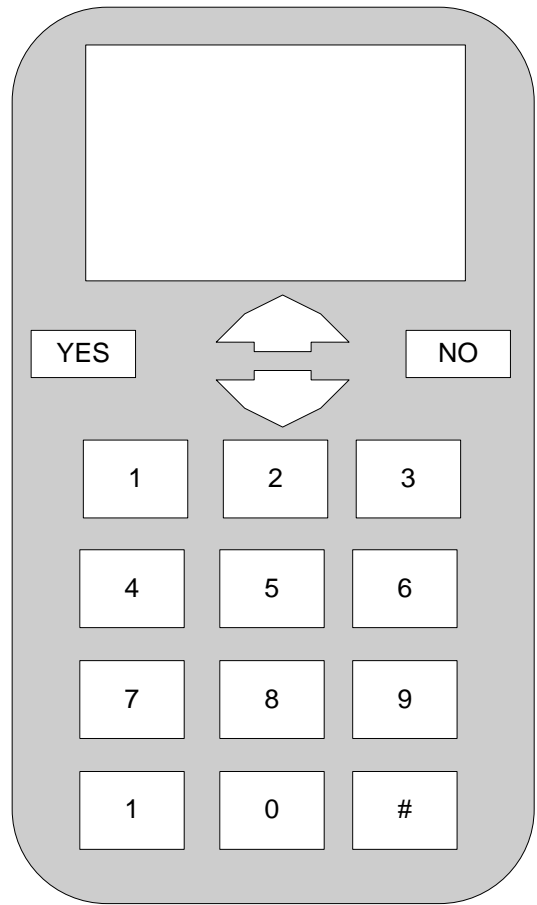

Figure 1

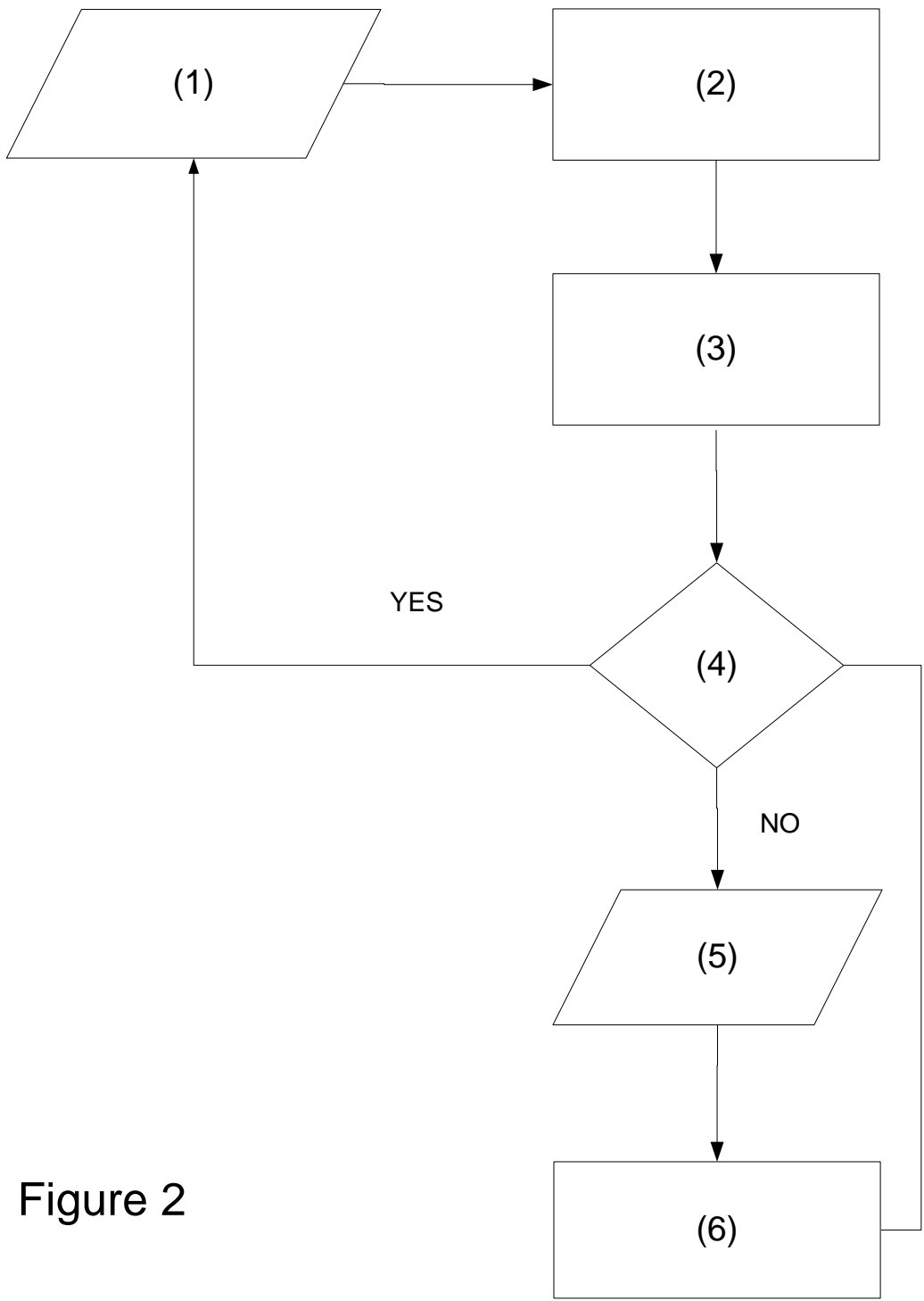

Figure 2

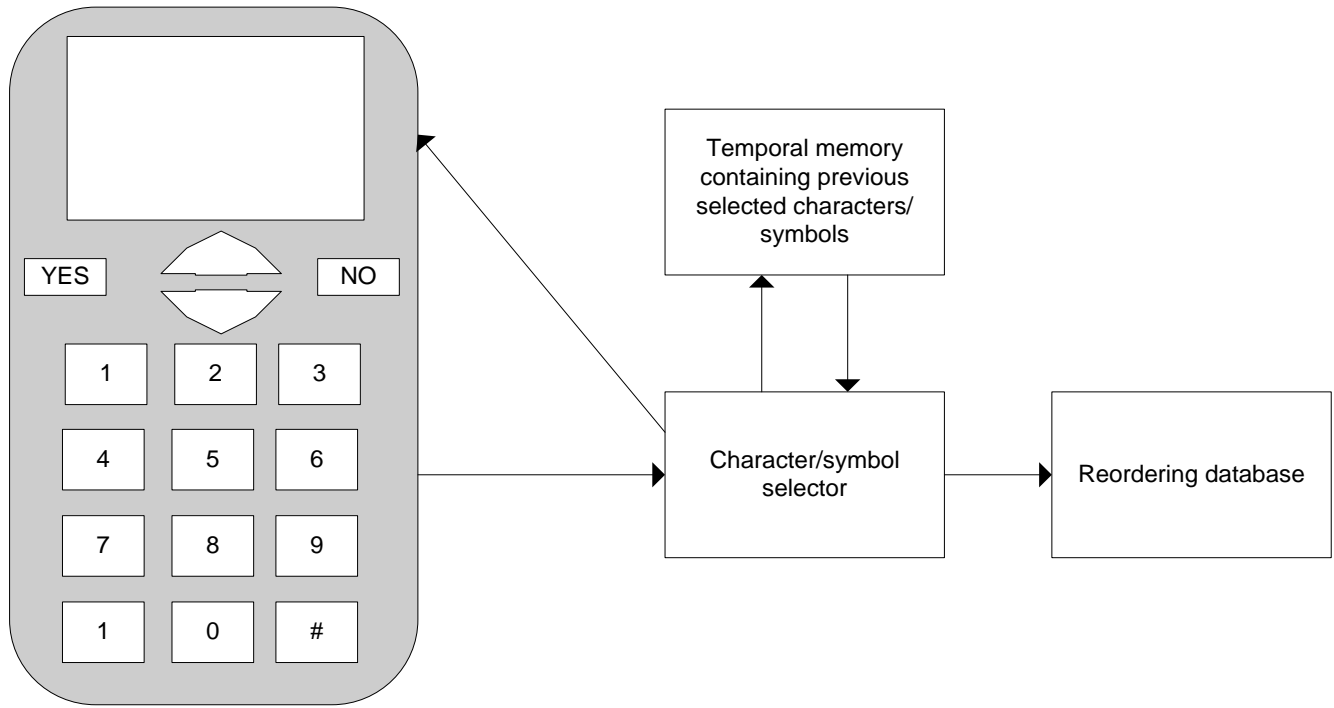

Figure 3

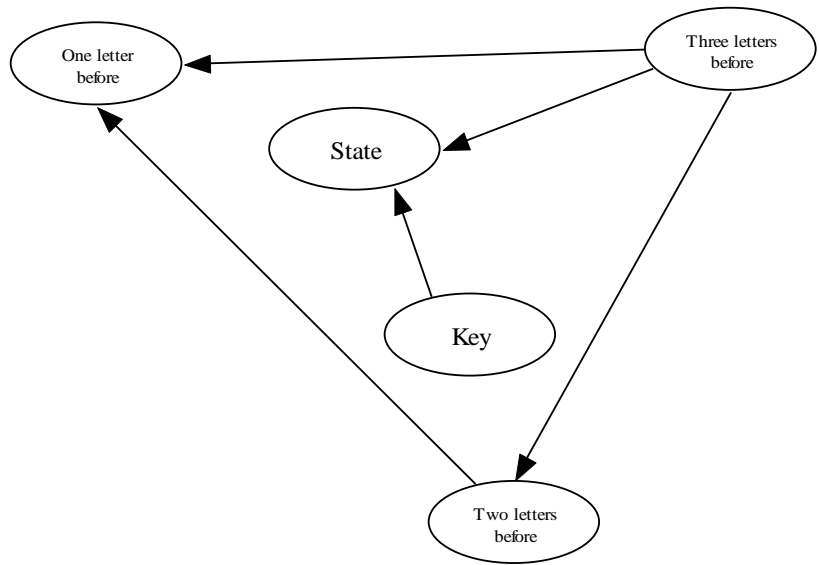

Figure 4

| PHRASES TO EVALUATE THE EFFICIENCY OF THE INVENTION (IN GREEK) |
|---|
| 1)ΜΠΟΡΕΙΣ ΝΑ ΠΕΡΑΣΕΙΣ ΑΠΟΨΕ ΑΠΟ ΤΟ ΣΠΙΤΙ ΚΑΤΑ ΤΙΣ ΔΕΚΑ ΝΑ ΜΙΛΗΣΟΥΜΕ ΓΙΑ ΤΟ ΤΑΞΙΔΙ; ΤΕΛΙΚΑ ΘΑ ΕΡΘΕΙ ΚΑΙ Ο ΜΑΝΩΛΗΣ ΜΑΖΙ ΜΑΣ ΕΛΠΙΖΩ ΝΑ ΜΗ ΣΕ ΠΕΙΡΑΖΕΙ |
| 2)ΑΥΡΙΟ ΔΕΝ ΘΑ ΑΝΕΒΩ ΠΑΝΕΠΙΣΤΗΜΙΟ ΓΙΑΤΙ ΜΟΥ ΕΤΥΧΕ ΚΑΤΙ ΠΟΛΥ ΣΟΒΑΡΟ ΣΧΕΤΙΚΑ ΜΕ ΤΗΝ ΑΝΤΙΓΟΝΗ ΙΣΩΣ ΜΠΟΡΕΣΩ ΝΑ ΣΟΥ ΚΑΝΩ ΕΝΑ ΒΙΑΣΤΙΚΟ ΤΗΛΕΦΩΝΗΜΑ ΚΑΤΑ ΤΟ ΒΡΑΔΑΚΙ |
| 3)ΚΑΛΑ ΕΧΑΣΕΣ ΦΟΒΕΡΟ ΣΚΗΝΙΚΟ ΜΕ ΤΟΝ ΑΛΕΚΟ ΗΡΘΕ ΤΡΕΧΟΝΤΑΣ ΑΠΟ ΤΟ ΓΥΜΝΑΣΤΗΡΙΟ ΚΑΙ ΣΤΟ ΔΡΟΜΟ ΕΧΑΣΕ ΤΟ ΠΟΡΤΟΦΟΛΙ ΤΟΥ ΜΑΖΙ ΜΕ ΟΛΑ ΤΑ ΧΑΡΤΙΑ ΚΑΙ ΤΑ ΛΕΦΤΑ ΜΕΤΑ ΧΑΜΟΣ |
| 4)ΕΒΡΙΖΕ ΦΩΝΑΖΕ ΧΤΥΠΟΥΣΕ ΣΑΝ ΤΡΕΛΛΟΣ ΤΟΝ ΤΟΙΧΟ ΓΕΝΙΚΑ ΞΕΣΗΚΩΣΕ ΣΤΟ ΠΟΔΙ ΟΛΗ ΤΗΝ ΥΨΗΛΑΝΤΟΥ |
| 5)ΜΗΠΩΣ ΕΧΕΙΣ ΚΑΘΟΛΟΥ ΚΑΙΡΟ ΝΑ ΔΕΙΣ ΓΙΑΤΙ Ο ΥΠΟΛΟΓΙΣΤΗΣ ΜΟΥ ΒΓΑΖΕΙ ΑΥΤΑ ΤΑ ΧΑΖΑ ΚΑΙ ΑΚΑΤΑΝΟΗΤΑ ΜΗΝΥΜΑΤΑ ΑΝ ΜΠΟΡΕΙΣ ΠΑΡΕ ΤΗΛΕΦΩΝΟ ΣΤΟ ΣΤΑΘΕΡΟ ΤΗΣ ΜΑΡΙΑΣ |
| 6)ΝΑ ΚΑΝΟΝΙΣΟΥΜΕ ΑΚΡΙΒΩΣ ΤΗΝ ΩΡΑ ΤΑ ΛΕΜΕ ΠΕΡΙΜΕΝΩ ΕΝΑΓΩΝΙΩΣ ΕΝΑ ΤΗΛΕΦΩΝΗΜΑ ΣΟΥ ΕΙΜΑΙ ΣΕ ΑΣΧΗΜΗ ΦΑΣΗ ΟΤΑΝ ΚΟΛΛΑΕΙ ΤΟ ΜΗΧΑΝΗΜΑ |
| 7)ΕΙΜΑΙ ΠΟΛΥ ΧΑΡΟΥΜΕΝΟΣ ΣΗΜΕΡΑ ΟΙ ΔΙΚΟΙ ΜΟΥ ΤΗΝ ΕΚΑΝΑΝ ΓΙΑ ΣΑΒΒΑΤΟΚΥΡΙΑΚΟ ΚΑΙ ΘΑ ΕΧΩ ΤΟ ΣΠΙΤΙ ΟΛΟ ΔΙΚΟ ΜΟΥ ΛΕΩ ΝΑ ΚΑΝΟΝΙΣΩ ΕΝΑ ΠΑΡΤΑΚΙ ΠΟΥ ΘΑ ΧΑΛΑΣΕΙ ΚΟΣΜΟ |
| 8)ΠΕΣ ΤΟ ΚΑΙ ΣΤΟ ΓΙΩΡΓΟ ΚΑΙ ΤΗ ΔΕΣΠΟΙΝΑ ΑΝ ΜΠΟΡΕΙΣ ΦΕΡΕ ΚΑΙ ΚΑΝΕΝΑ ΔΙΣΚΑΚΙ ΜΑΖΙ ΕΝΤΑΞΕΙ |
| 9)ΕΛΑ ΦΙΛΕ ΔΙΑΒΑΖΕΙΣ ΕΜΕΙΣ ΚΑΝΟΥΜΕ ΜΠΑΝΑΚΙ ΚΑΙ ΒΛΕΠΩ ΤΟ ΝΟΤΗ ΝΑ ΠΙΝΕΙ ΦΡΑΠΕΔΑΚΙ ΟΙ ΤΟΥΡΙΣΤΡΙΕΣ ΠΟΛΥ ΚΑΛΕΣ ΚΑΛΟ ΚΟΥΡΑΓΙΟ |
| 10)ΚΑΛΗΣΠΕΡΑ ΝΙΚΟΛΑ Ο ΜΗΤΣΟΣ ΕΙΜΑΙ ΕΧΕΙΣ ΤΟ ΦΟΟΥΝ ΤΟΥ ΛΕΥΤΕΡΗ |

Figure 5

| Phrase | # words | Characters | iPRETI | STEM | % improvement | Single erors | Double errors |
|---|---|---|---|---|---|---|---|
| 1 | 28 | 143 | 158 | 256 | 38.3% | 9.1% | 0.7% |
| 2 | 29 | 160 | 169 | 268 | 36.9% | 5.6% | 0.0% |
| 3 | 29 | 158 | 179 | 279 | 35.8% | 8.2% | 2.5% |
| 4 | 14 | 87 | 102 | 142 | 28.2% | 8.0% | 4.6% |
| 5 | 25 | 150 | 165 | 265 | 37.7% | 8.7% | 0.7% |
| 6 | 20 | 122 | 134 | 218 | 38.5% | 6.6% | 1.6% |
| 7 | 28 | 154 | 169 | 276 | 38.8% | 7.1% | 1.3% |
| 8 | 16 | 86 | 92 | 150 | 38.7% | 4.7% | 1.2% |
| 9 | 19 | 117 | 129 | 213 | 39.4% | 8.5% | 0.9% |
| 10 | 10 | 58 | 63 | 107 | 41.1% | 8.6% | 0.0% |
| **Total** | **218** | **1235** | **1360** | **2174** | **37.4%** | **7.5%** | **1.3%** |

Figure 6